\documentclass[twocolumn]{aastex63}

\usepackage{amsmath}
\usepackage{graphicx}
\usepackage{natbib}
\usepackage{lineno}

\shorttitle{XX Oph \& AS 325}
\shortauthors{Howell et al.}

\begin{document}

\title{Inside the Iron Curtain: A Long Term Look at the Iron Stars XX Oph and AS 325} 


\correspondingauthor{Steve B. Howell}
\email{steve.b.howell@nasa.gov}

\author[0000-0002-2532-2853]{Steve~B.~Howell}
\affiliation{NASA Ames Research Center, 
Moffett Field, CA 94035 USA}

\author[0000-0002-4641-2532]
{Venu~M.~Kalari}
\affiliation{Gemini Observatory/NSF's NOIRLab, Casilla 603, La Serena, Chile}

\author[0000-0003-1120-5178]
{Andy~Adamson}
\affiliation{Gemini Observatory/NSF NOIRLab, 670 N. A’ohoku Place, Hilo, HI, 96720, USA}

\author[0000-0002-0885-7215]
{Mark~Everett}
\affiliation{WIYN Observatory/NSF’s NOIRLab, 950 N. Cherry Avenue, Tucson, AZ 85719, USA}

\begin{abstract}
The Iron Stars XX Oph and AS 325, both binaries consisting of a Be + late K,M II star present remarkable optical spectra. Low velocity, dense winds from the late type star collide with high velocity, optically thin material expelled from the hot Be star producing a plethora of emission lines and complex P Cygni absorption profiles.
The members of the American Association of Variable Star Observers have faithfully observed XX Oph for 85 years and AS 325 for 32 years. This long term photometric monitoring has revealed AS 325 to be an eclipsing system while XX Oph shows a complex light curve behavior. We present archival and recent photometry, new high-resolution optical imaging, and new high-resolution optical spectroscopy of the two stars. 
The orbital period of AS 325 is refined as 512.943 days and the stellar components are Be+K2.5 II. XX Oph is shown to be non-eclipsing and consists of Be+M6II stars.
\end{abstract}



\section{Introduction} 

One of the most fascinating stages in stellar evolution is that of the asymptotic giant branch (AGB) and the Semi-Regular variable stars.
Place one such star in a binary system, and the photometric and spectral properties yield extraordinary, complex results. Two such binary stars, AS 325 and XX Oph, epitomize the case. 

XX Oph (HD 161114, Merrill's Iron Star, d=2143 pc) was discovered by Williamina Fleming\footnote{https://en.wikipedia.org/wiki/Williamina\_Fleming} \citep{Fleming1908HarCi.143....1F}. Since then, numerous studies have revealed the confusing and fantastic properties of this star. \citet{Prager1940BHarO.912...17P} presented the first long-term light curve of XX Oph revealing its non-constant behavior, having a mean maximum near m$_{pg}$=9.6 with occasional, irregular, rapid drops to m$_{pg}$=10.2-11. Spectrally, XX Oph shows a wondrous, changing, complex spectrum essentially unequaled in astronomy. The forest of Fe II emission lines across the optical spectrum gave rise to XX Oph being dubbed the Iron Star. \citet{Merrill1924PASP...36..225M,Merrill1932ApJ....75..133M,Merrill1951ApJ...114...37M,Merrill1961ApJ...133..503M} provides a 30-year baseline study covering the many facets of spectral features in this star. Other studies \citep[e.g.,][]{deWinter1990Ap&SS.166...99D,Cool2005PASP..117..462C,Howell2009PASP..121...16H} have been undertaken as well, including three single epoch high-resolution spectral observations \citep{Goswami2001BASI...29..295G,tarasoc2006ASPC..355..297T,TOM2010AJ....140.1758T}.  The consensus of all of the spectral studies is that the numerous bright metallic emission lines due to Fe, Ti, and Cr exhibit only small velocity and shape changes, whereas the P Cygni absorption lines of hydrogen, helium, sodium, and calcium, appear to behave erratically. XX Oph is believed to be a Be (V or III) + M6 III non-eclipsing binary of unknown orbital period \citep{lockwood1975ApJ...195..385L,Evans1993A&A...267..161E,Cool2005PASP..117..462C}.

\begin{figure*}
    \centering
\includegraphics[width=0.9\textwidth]{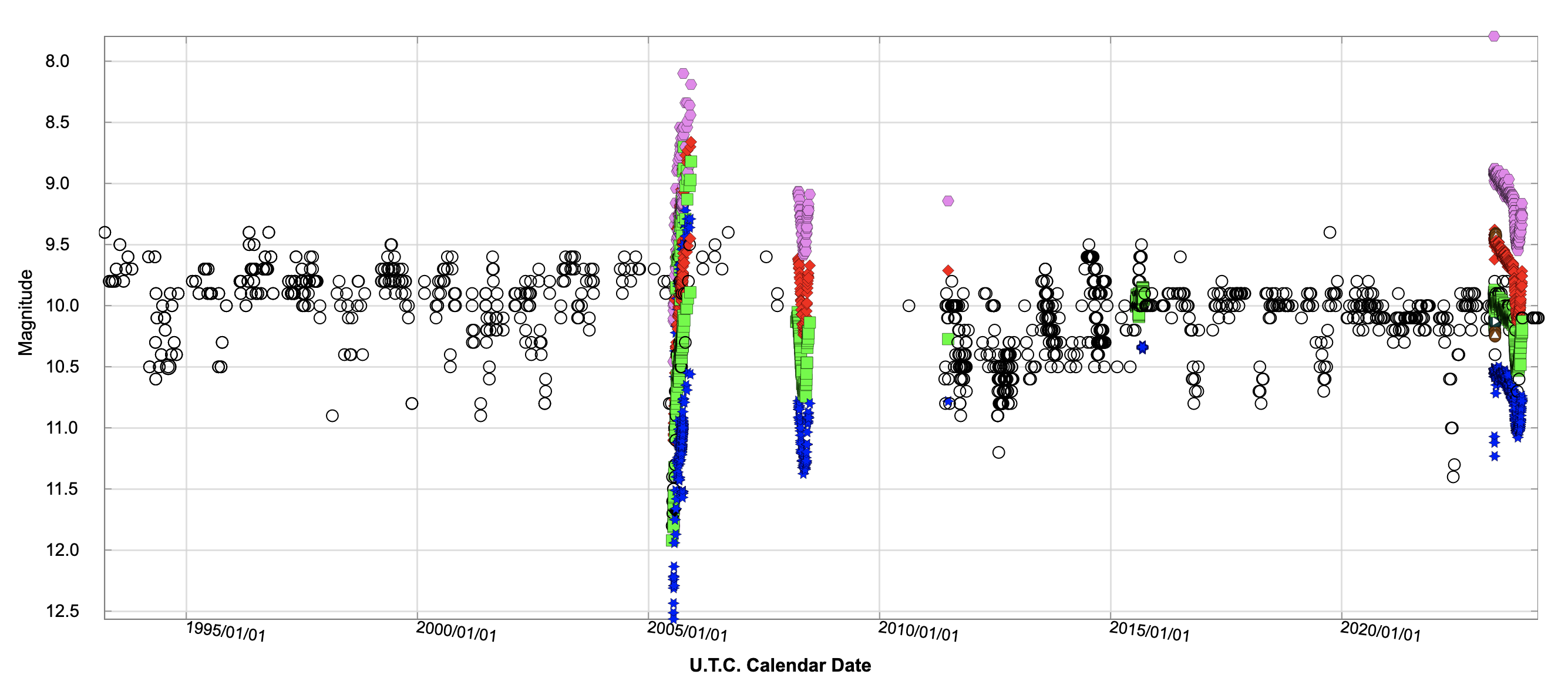}
\caption{32 years of AAVSO photometry for AS 325. Black symbols are visual points, while colored points are B, V, R, I CCD observations. The three multi-color eclipse campaigns are shown in more detail in Figure 2. Data obtained from https://www.aavso.org/LCGv2/.}
\end{figure*}

AS 325 (CD -26$^o$13521, V5569 Sgr, d=5339 pc) was first listed as a star with a bright H$\alpha$ emission line in 1950 \citep{AS-Ha1950ApJ...112...72M}. In that paper, a note attached to the stars' entry (No. 325) said, ``The type is Fe. The star deserves further observation." \citet{Bopp1989PASP..101..981B} provide a historical summary of AS325 covering its discovery up to the time of their paper in which AS 325 was declared the second only known Iron Star.
AS325 exhibits an optical spectrum that contains a surplus of emission lines including hydrogen, sodium, calcium, titanium, and chromium and, most abundantly, iron. Helium is not present \citep{Bopp1989PASP..101..981B,Cool2005PASP..117..462C,Howell2009PASP..121...16H}. Each emission line often has an associated P Cygni profile and the star has been shown to be an eclipsing binary system consisting of a Be V + K2.5-M5 II variable in a 513 day orbit \citep{Otero2005IBVS.5608....1O, Howell2009PASP..121...16H}. AS 325 does not have as long and rich an astronomical history as XX Oph, but does have a 32 year photometric light curve showing over 20 stellar eclipses and a nearly constant out of eclipse visual magnitude of 10.0.

The binary stars XX Oph and AS 325 present remarkable optical spectra. They consist of ever changing absorption lines caused by mass loss episodes and winds plus an array of metal emission lines due to shocks within low density colliding winds in an extended halo. Their spectral nature is distinctly different at different epochs. The observed spectral changes and mass loss events are consistent with the behavior of yellow hypergiants \citep{Hyoer1998A&ARv...8..145D}. 

Both of the iron stars, XX Oph and AS 352, reside in dense regions of the Milky Way. AS 325 is located in Sagittarius while XX Oph is located in Serpens Cauda (not current day Ophiuchus)\footnote{In figurative representations, the body of the Serpent is represented as passing behind Ophiuchus, the Serpent Bearer, and separated into Serpens Caput (The Head) and  Serpens Cauda (The Tail).} The locations within our Galaxy of the two stars have been discussed in some detail in \citet{Cool2005PASP..117..462C,Howell2009PASP..121...16H}. 

In this paper, we review many years of photometric and spectral observations for the stars AS 325 and XX Oph. We also present recent photometry, new high-resolution optical images using NESSI on WIYN and Zorro on Gemini South, and new high-resolution optical spectroscopy using GHOST at Gemini South.
We refine the orbital period of AS 325 and confirm the non-eclipsing nature of XX Oph.

\section{Observations}
\subsection{AS 325}
\subsubsection{Long-Term photometry}

AS 352 was brought to the attention of astronomers in 1989 \citep{Bopp1989PASP..101..981B} and
was first observed by the American Association of Variable Star Observers (AAVSO) on 01 March 1993. The AAVSO visual and multi-color observations continue until the present time, covering a time period of 11,325 days and spanning 22 eclipses of AS 325.
Figure 1 presents the complete set of AAVSO visual observations tirelessly gathered by their intrepid observers. We note a number of obvious features in this long-term light curve. During the years 1994 to 2009, AS 325 had a mean out-of-eclipse visual magnitude of 9.7. During the years 2010 to 2015, the light curve is somewhat chaotic and has a mean magnitude near 10.4 and finally, in the most recent years, AS 325 has a mean magnitude near 10. The time span of 2010 to 2015 is also a time when the eclipse profiles are difficult to measure or even discern based on these data. Table 1 assigns each eclipse (or eclipse time) a number and presents the best estimate for mid-eclipse times from the measured AAVSO light curve using Argelanders method \citep{1844scja.book..122A}. Eclipse durations are estimated as best as possible from the light curve observations, attempting to span the eclipse start to end with ``zero" being set to the local out-of-eclipse light curve level. The three multi-color AAVSO ``campaign" eclipses (eclipses 9,10,22) are far better sampled in time, thus these eclipse parameters have less uncertainty, $\pm$1-2 days. Some of the eclipses are fairly uncertain due to sparse data and some are completely unmeasurable due to the fact that AS 325 was near the sun (Nov to Feb) and no data was available, the data was very sparse, or the light curve was quite chaotic. The more uncertain eclipse measurements are indicated in Table 1 with a ``:" attached to the eclipse number. We will make use of these eclipses for an orbital period determination in \S3.1.2.

To illustrate the variety of 
eclipses that AS 325 has, we see in Figure 2, the three best studied eclipses from the AAVSO multi-color campaigns (numbers 9, 10, and 22). Eclipse 9 was 2 magnitudes deep and had a duration of 227 days, eclipse 10 has a depth of only 0.6 magnitudes and lasted 92 days, while eclipse 22 had a depth of 0.5 magnitudes and lasted a brief 60 days. The latter two eclipses are both shallow and short in comparison to eclipse 9.
At times, AS 325 also shows consecutive eclipses that are very similar in depth and duration, eclipses 17-19 for example. 

\begin{figure*}
    \centering
    \includegraphics[width=0.8\textwidth]{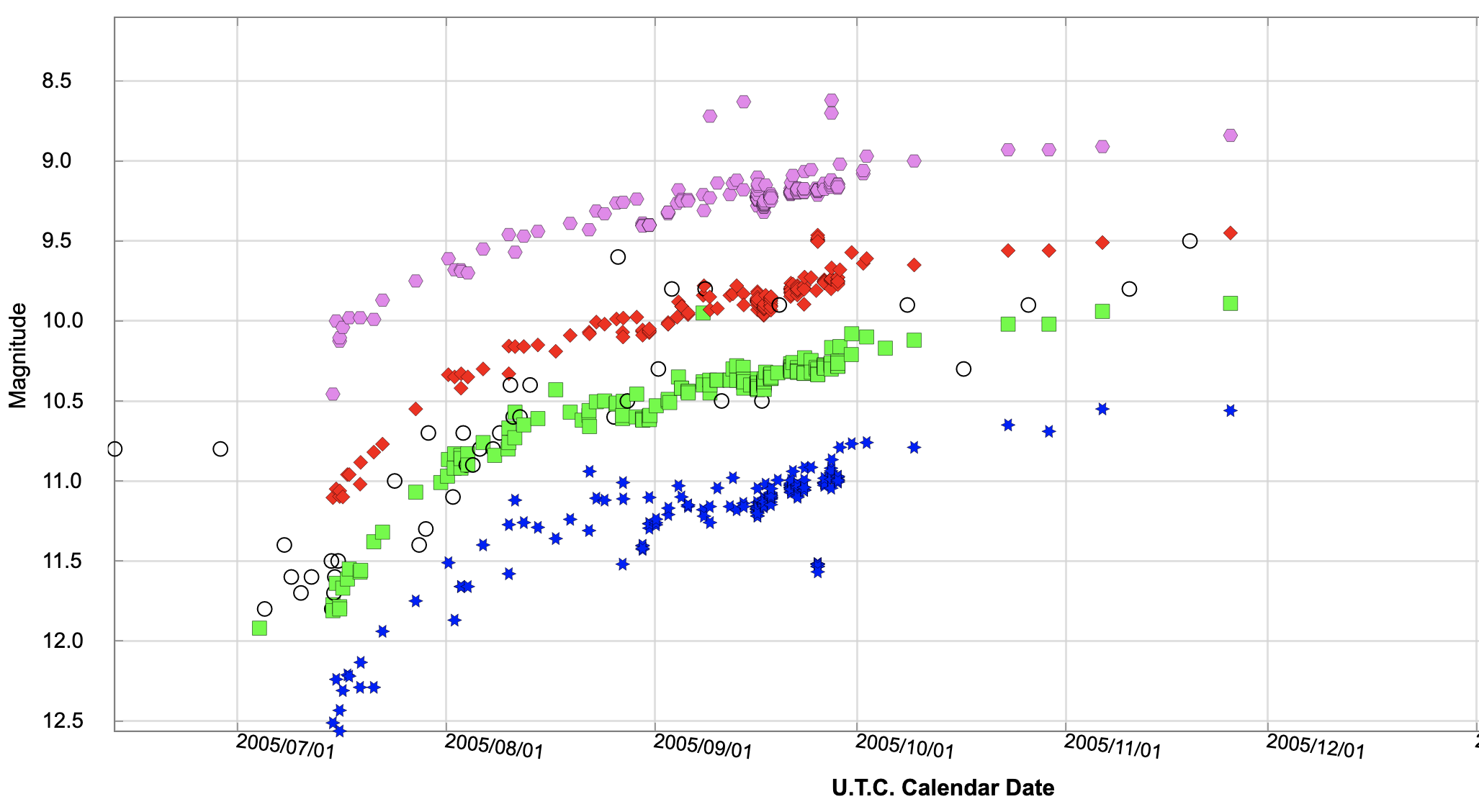}
    \includegraphics[width=0.8\textwidth]{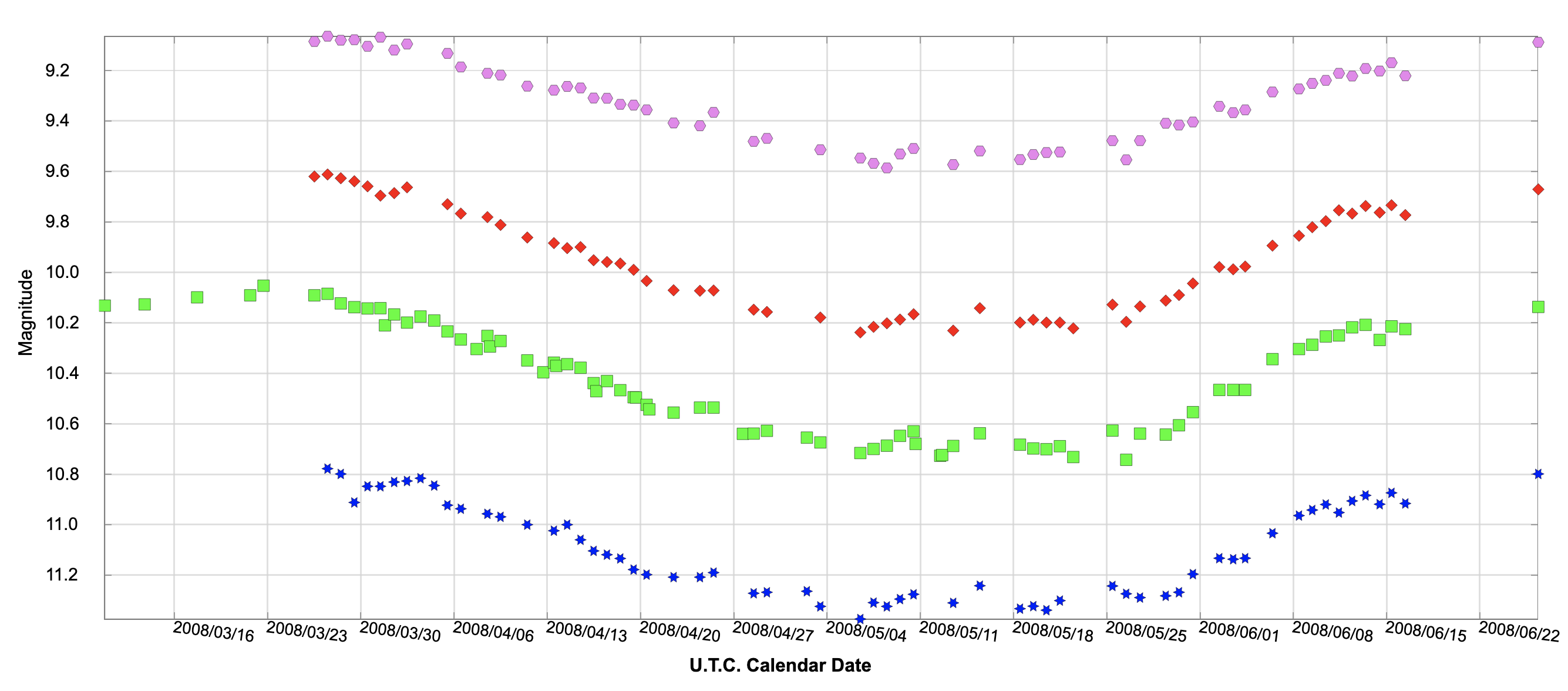}
    \includegraphics[width=0.8\textwidth]{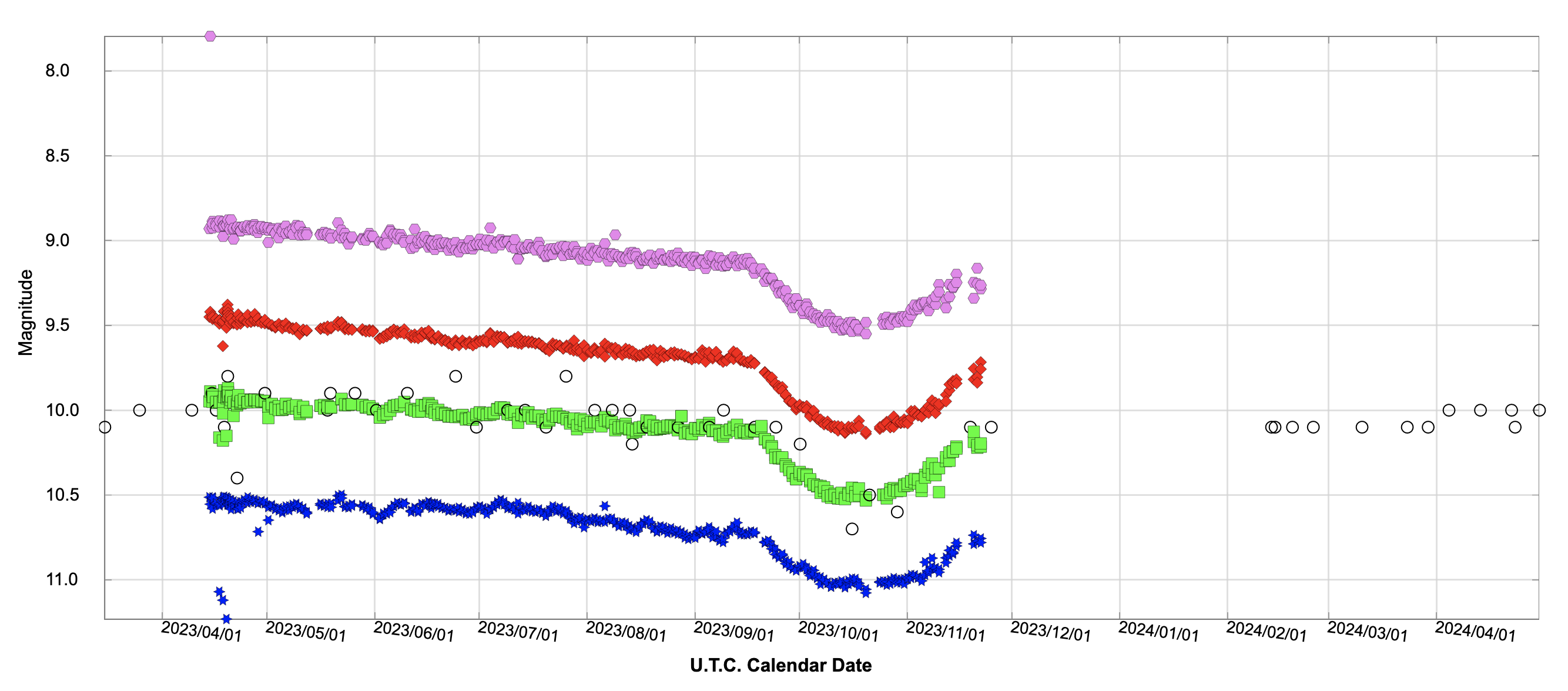}
    
    \caption{The three multi-color AAVSO eclipse campaigns of AS 325. From top to bottom: Eclipses 9, 11, and 22 are shown in detail with depths of 2.0, 0.6, and 0.5 magnitudes in V and durations of 227, 92, and 60 days respectively. Eclipse 9 is shown in full in \citet{Howell2009PASP..121...16H}. Blue, Green, Red, and Pink points are Johnson B, V, R, I filter CCD observations respectively and black symbols are visual measurements. The GHOST high resolution spectrum was obtained on 17 April 2023, just at the start of the eclipse 22 multi-color campaign. Note that the x-axis time scales are unequal for the three eclipses shown here.
    Data obtained from https://www.aavso.org/LCGv2/.}
\end{figure*}  

\begin{figure*}
    \centering
\includegraphics[width=0.45\textwidth]{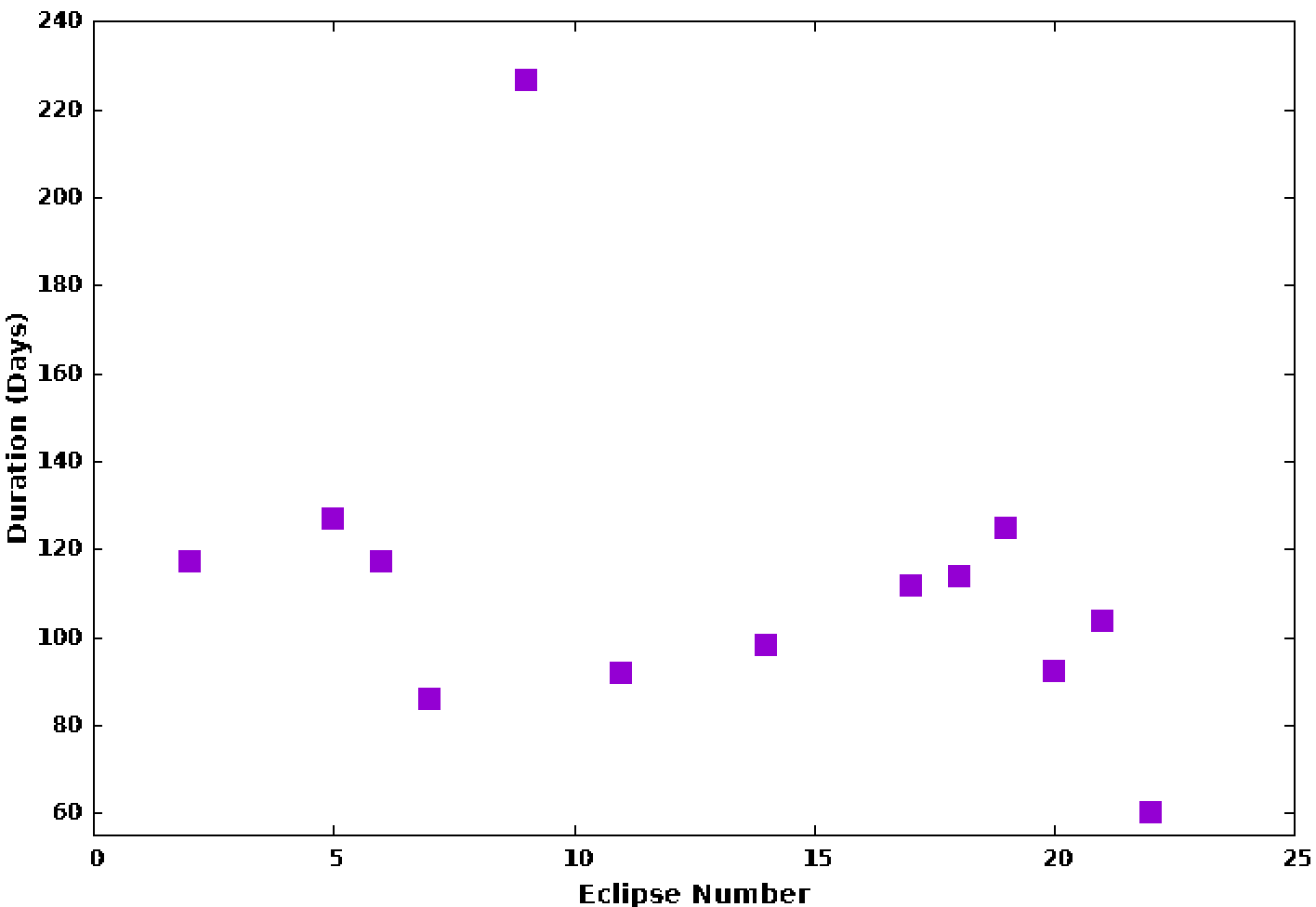}
\includegraphics[width=0.45\textwidth]{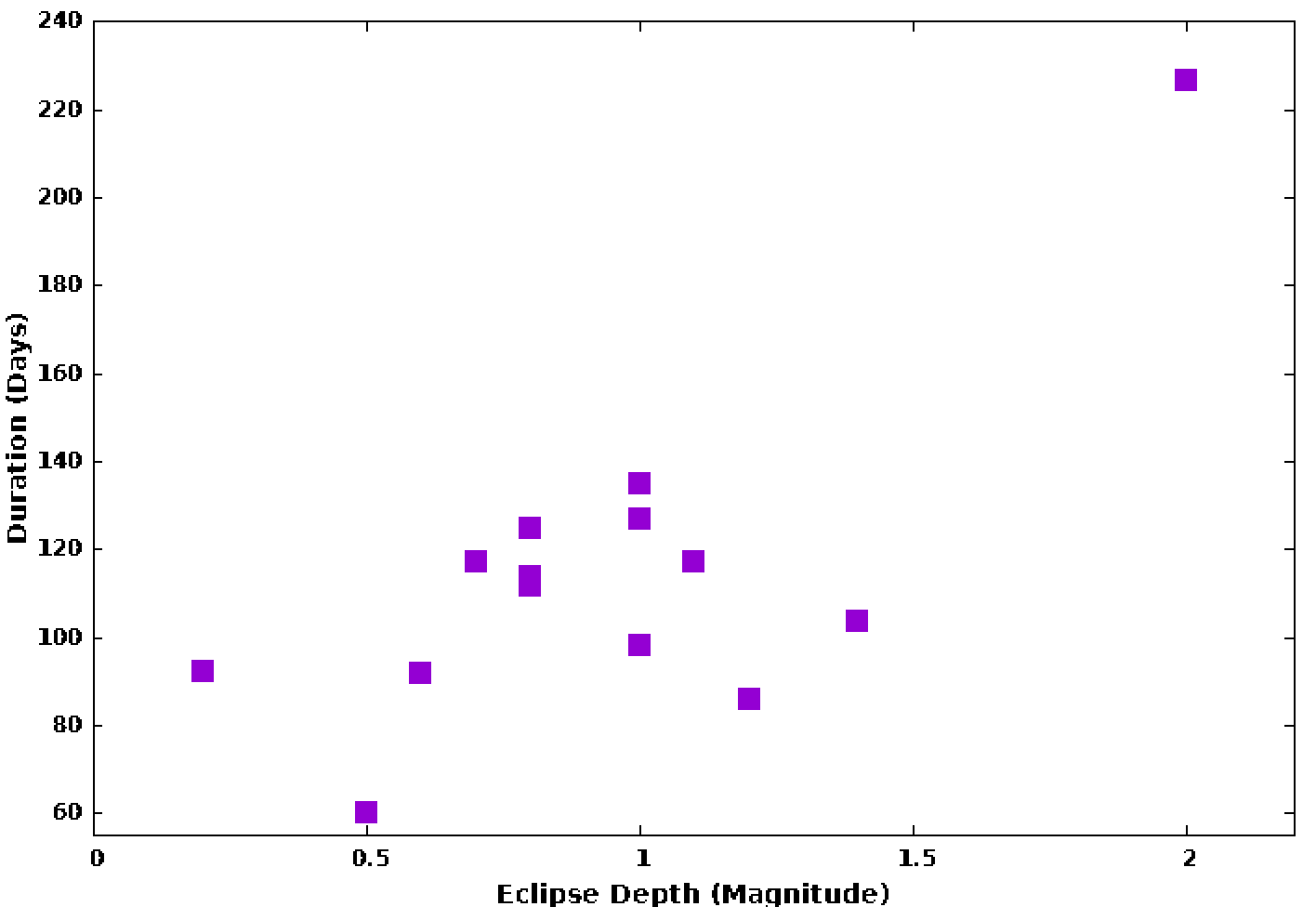}
    \caption{Left:Total eclipse duration in days for those eclipses that could be well estimated over the 32 year time period covered by the AAVSO observations. Eclipse number is listed in Table 1. 
    Right: Eclipse depth in magnitudes vs. duration in days. Note that eclipse 9 has the longest duration by far covering nearly one-half of the entire orbital period at a depth of 2 magnitudes. Most eclipses have durations near 100 days and depths near 1 magnitude.}
\end{figure*}   

Noting that shallow eclipses tend to have shorter durations, Figure 3 shows the duration of the eclipse over time as indicated by the eclipse number. 
Figure 3 also shows the relation between eclipse duration and eclipse depth, longer durations being deeper. \citet{Howell2009PASP..121...16H} related the trends seen in eclipse duration and depth to the variable nature and radius of the red companion. 
Some of the variability in the eclipse depths could be caused by an inclined orbit, which may explain the apparent total lack of eclipses at times. Alternating primary and secondary eclipses can be ruled out using the spectroscopic evidence presented in \citep{Howell2009PASP..121...16H, Cool2005PASP..117..462C}.
There may be a hint of periodic behavior in the duration of the eclipse
on the order of $\sim$10 years (the red star pulsation period?).
The eclipse depth seems to remain near 1.0 magnitude for most of the time, with a few exceptions (Table 1). Figures 1 and 2 show the effect on the eclipse shape and depth, and Table 1 shows that some eclipses seem to be completely missed, possibly indicative of a critical orbital inclination allowing no eclipses to occur when the red star is at its smallest size.

\begin{deluxetable*}{ccccc}
\tablewidth{0pt}
\tablecaption{Eclipses of AS325 \label{tab:eclipse}}
\tablehead{
\colhead{Eclipse No.} & \colhead{Date} & \colhead{Mid-Time (JD)} & \colhead{Depth (Visual mag)} & \colhead{Duration(days)}
}
\startdata
1: & June 1994 & 2449504.93 & 0.7 & $<$233 \\
2: & Oct 1995 & 2449992.92 & 0.7 & 116.96 \\
3 & Sparse/no data & & & \\
4: & Feb 1998 & 2450872.174 & 1.0 & $<$135 \\
5: & Nov 1999 & 2451500.8903 & 1.0 & 126.76 \\
6: & May 2001 & 2452011.236 & 1.1 & 117.04 \\
7 & Oct 2002 & 2452524.9483 & 1.2 & 85.88 \\
8 & Sparse/no data & & & \\
9 & July 2005 & 2453550.7458 & 2.0 & 226.73 \\
10 & Sparse/no data & & & \\
11 & May 2008 & 2454576.94309 & 0.6 & 91.83 \\
12 & Sparse/no data & & & \\
13 & Sparse/no data & & & \\
14: & July 2012 & 2456163.65972 & 1.0 & 98 \\
15: & Complex/Feb 2014& 2456703.01389 & 1.0 & $<$200 \\
16: & Complex/June 2015 & 2457174.91736 & 0.5 & $<$ 210\\
17 & Oct 2016 & 2457689.5333 & 0.8 & 111.83 \\
18 & Mar 2018 & 2458200.95139 & 0.8 & 113.72 \\
19 & Aug 2019 & 2458714.61875 & 0.8 & 124.68 \\
20: & Feb 2021 & 2459253.00139 & 0.2 & 92 \\
21 & May 2022 & 2459726.73819 & 1.4 & 103.75 \\
22 & Oct 2023 & 2460238.55625 & 0.5 & 60.19 \\
\enddata
\tablecomments{Eclipses with a ``:" contain very uncertain values due to sparse sampling or chaotic light curve behavior. Uncertainties in these eclipse mid-times are 5-10 days. Eclipses 15 and 16 occurred during a very complex portion of the AS 325 light curve as such, their duration is highly uncertain.}
\end{deluxetable*}

\subsubsection{High-Resolution Imaging}

To provide a higher-resolution image of the local scene near AS 325 and to search for any possible confounding close companions, we obtained high-resolution optical speckle images
using NESSI mounted on the WIYN 3.5-m telescope located at Kitt Peak National Observatory.  NESSI \citep{2018PASP..130e4502S}
provides simultaneous speckle imaging in two optical bands with output data products including robust 5$\sigma$ magnitude contrast limits.

AS 325 was observed on 02 July 2018 in which we obtained five sets of 1000$\times$0.04 sec exposures and subjected them to Fourier analysis in our standard reduction pipeline \citep{2011AJ....142...19H}. Figure 4 shows the 5$\sigma$ contrast curve obtained for AS 325 observed at 832 nm. 
No close stellar companion to the binary was detected in the high-resolution imaging within the magnitude contrast achieved nor within the angular limits from the diffraction limit (64 mas) out to 1.2 arcsec, corresponding to spatial limits of 341 to 16,000 AU at the distance of AS 325 (d=5339 pc). At the distance of AS 325, we did not, of course, expect to be able to separate the two stellar components.

\begin{figure}
    \centering
\includegraphics[width=0.45\textwidth]{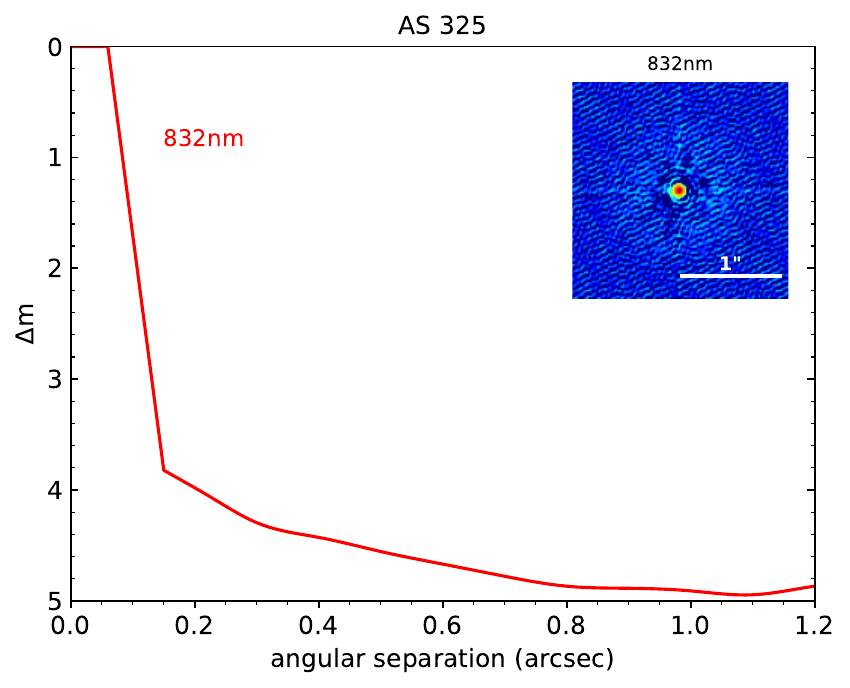}
\caption{WIYN NESSI high-resolution optical speckle imaging result for the star AS 325. The red line shows the 5$\sigma$ magnitude contrast result and the inset is the speckle reconstructed image at 832 nm. No close stellar companions or extended nebulosity were detected to within 4-5 magnitudes of AS 325 from the diffraction limit to a radius of 1.2 arcsec.}
\end{figure}   

\subsubsection{Historical Spectroscopy}

AS 325, a system showing periodic eclipses (P=513 days), allows for some ease of interpretation as the orbital phase and nature of the binary, a Be + K2.5III system \citep{Cool2005PASP..117..462C,Howell2009PASP..121...16H} is well established. Spectral discussion in \citet{Bopp1989PASP..101..981B} suggested that the star is similar to Merrill's Iron Star, XX Oph. The general optical spectral appearance shows bright emission lines of hydrogen, Fe II, Ti, and other metals \citep{Cool2005PASP..117..462C}
but no He is present.
Absorption lines proceed from weak to strong as they move from low to higher blue shifted P Cygni profiles and are seen in essentially every spectral line.
Using a number of years of low resolution spectroscopy in connection with the AAVSO long term light curve of AS 325, \citet{Howell2009PASP..121...16H} produced a model for AS 325 in which the red secondary star varies between a K2.5 II and a M5 II star, changing its radius from $\sim200 $R$_{\odot}$ to $\sim500 $R$_{\odot}$ causing eclipse depth and width changes (See Table 1) as well as spectral changes (P Cygni profiles TiO bands, and Ca H\&K absorption) over many hundreds of days.  A detailed line list for optical, UV, and IR spectra for AS 325 are presented in \citet{Cool2005PASP..117..462C,Howell2009PASP..121...16H}.

\subsubsection{GHOST Spectroscopy}

High-resolution optical spectra were obtained for AS 325 using the GHOST (Gemini High Resolution Spectrograph; \citealt{2024AJ....168..208K, 2024PASP..136c5001M}) spectrograph, mounted on the Gemini South telescope on Cerro Pach{\'o}n, Chile. Spectra were obtained on the 17$^{th}$ April, 2023 (JD=2460052.3), covering 347-1060\,nm (although S/N to fully identify lines is limited to around 370-1000\,nm) at a resolution of 56,000. During the observations, the seeing hovered around 1", with thin cirrus cloud cover present. In the blue camera, with the red wavelength cutoff of 530\,nm, a single 300s exposure was taken, and in the red camera three sequential 60s exposures were taken for each star. 

Data reduction was performed using the Gemini DRAGONS data reduction pipeline, including bias subtraction, flat fielding and wavelength calibration \citep{DRAGONS2019ASPC..523..321L}.
The spectra were sky subtracted and extracted using variance weighting. The final extracted blue spectrum has continuum signal to noise ratios  between 20 (at 380\,nm) to 80 (at 510\,nm). Single red spectra achieved signal to noise ratio between 50 (at 550, and 950\,nm), peaking at 75 (around 800\,nm).  Flux calibration was performed using standard star observations obtained the same night as the science observations. Due to the many emission lines present, the order-merged spectra do not always cleanly merge at the order end, even after correcting the fit. 

Identification and estimation of the line properties was performed using normalized spectra before flux calibration. We discuss in more detail the emission lines, absorption lines, and diffuse interstellar bands (DIBs)below.

\begin{figure*}
\includegraphics[width=1\textwidth]{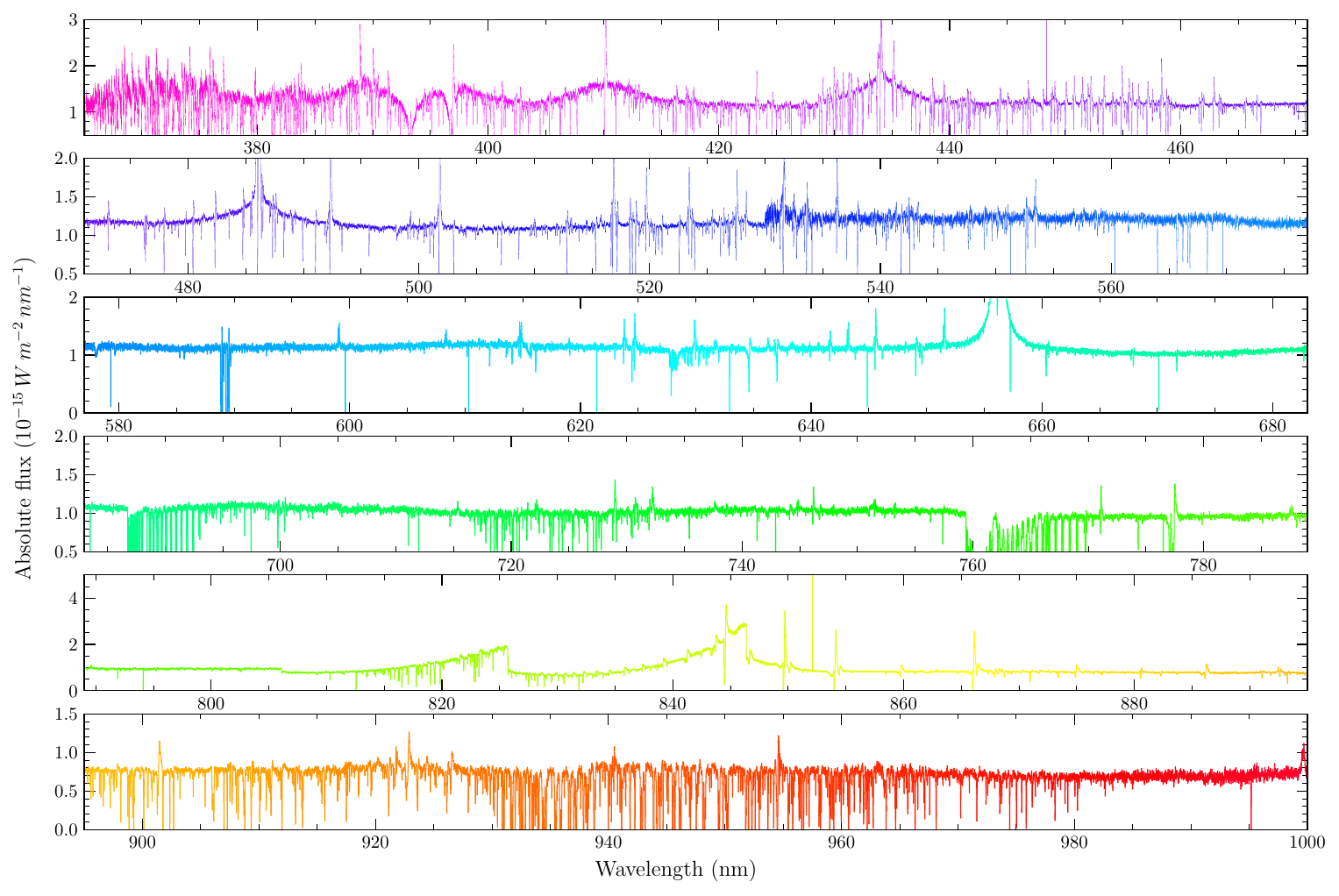}
\caption{Flux calibrated high resolution GHOST spectrum of AS 325 obtained on 17 April 2023. The various spectral orders are color coded for convenience showing the complex spectrum from 370-1000 nm. \label{aspectra}}
\end{figure*}

The flux-calibrated extracted GHOST spectra of AS 325 is shown in Fig.\,\ref{aspectra}.
The GHOST spectrum of AS 325 was obtained near the beginning of the AAVSO multi-color photometry campaign for eclipse 22 (See Table 1). These are the first high resolution spectral observations of this star. Looking at the bottom panel of Figure 2, we note that the GHOST observations occurred just a few months before mid-eclipse (JD=2460052.3) at orbital phase 0.672 (using the ephemeris given below). At this phase, the Be star is moving around toward the front of the red star from our perspective, heading toward eclipse 22 \citep[See Figure 10 in][]{Howell2009PASP..121...16H}. 

\begin{figure*}
\centering
\includegraphics[width=0.45\textwidth]{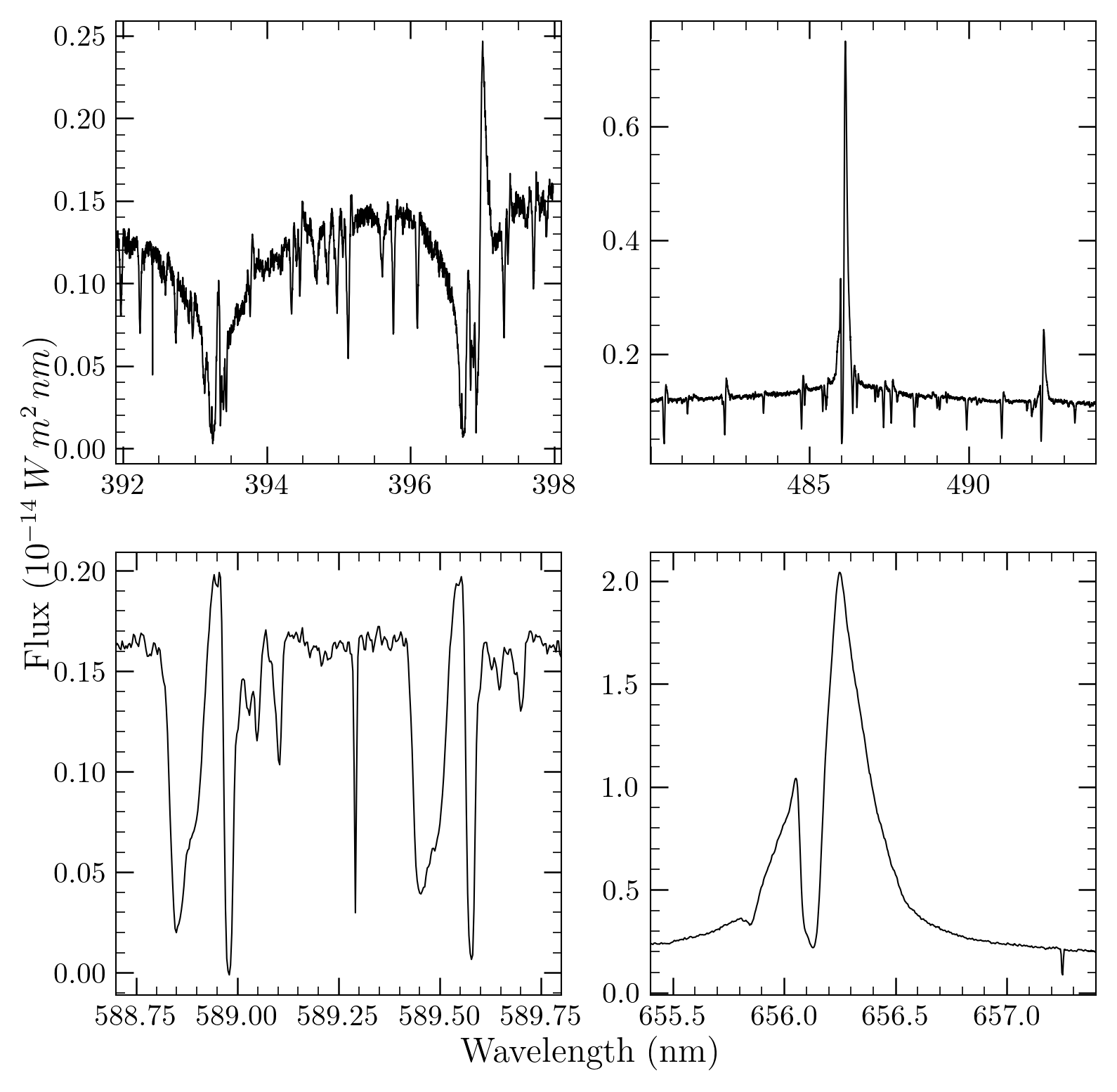}
\includegraphics[width=0.45\textwidth]{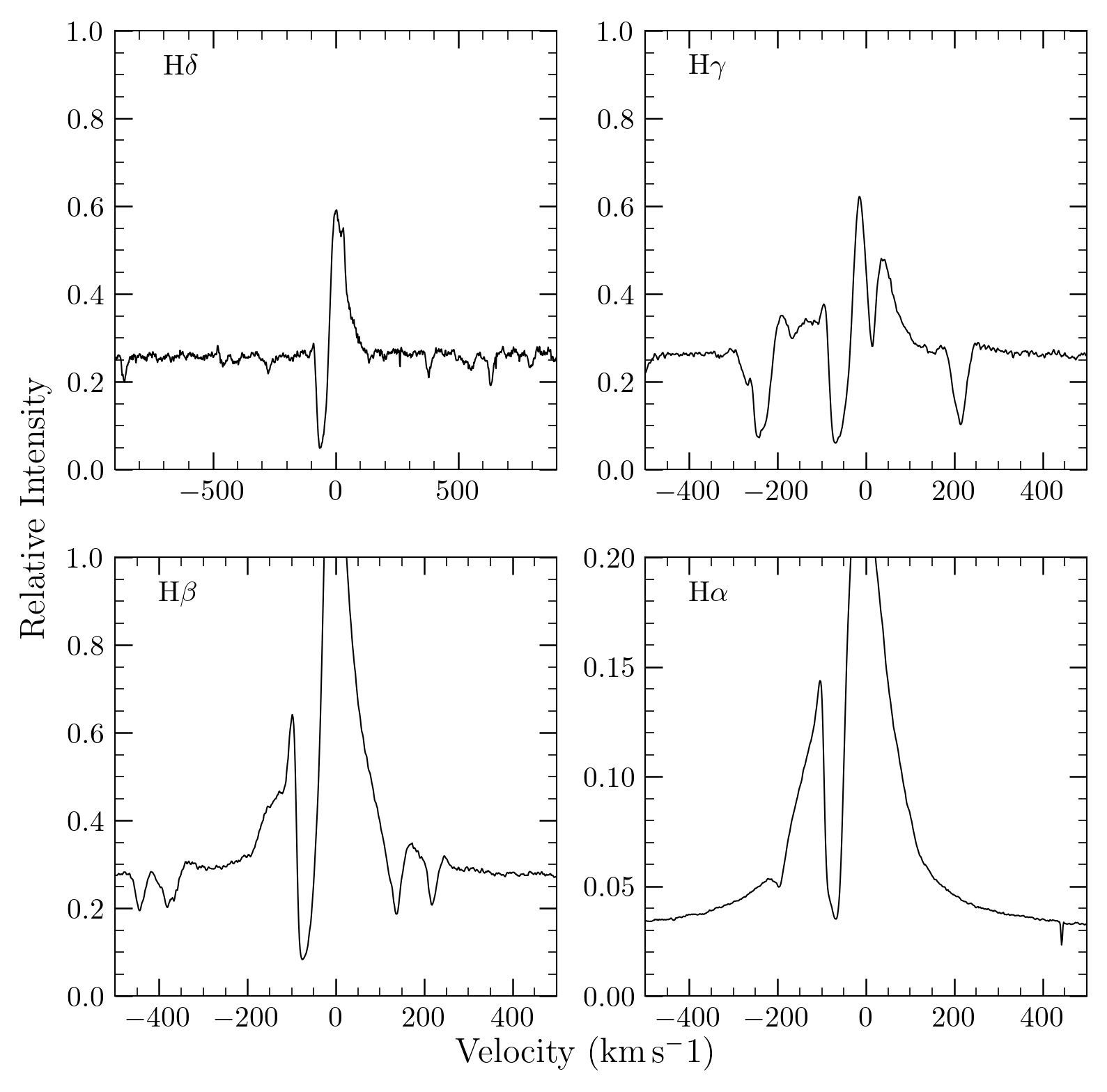}
\caption{Zoomed in selected spectral regions of interest for AS 325 covering Ca II H\&K, H$\beta$, NaI D, H$\alpha$ (Panel a), with select Balmer lines shown in (Panel b).
\label{azooms}}
\centering
\end{figure*}

In Fig.\,\ref{azooms}
we show detailed views of highlighted regions in the GHOST spectrum of AS 325 as well as the strongest Balmer lines. We note the Ca II H\&K absorption from the red star with weak emission cores, due to stellar activity or the Be star. H$\beta$ and H$\alpha$ are strongly in emission but both show blue shifted P Cygni absorptions. These absorptions are seen to move blueward over time reaching velocities of $-$100 to $-$200 km\,sec$^{-1}$ as the dense episodic red star wind expands \citep{Cool2005PASP..117..462C}. The NaI D lines are complex showing blue shifted absorptions at the same velocity at the hydrogen lines and red shifted absorption components possibly due to line of sight/local ISM material. Table 2 lists the measured velocity values for these lines.

\begin{deluxetable}{ccc}
\tablewidth{0pt}
\tablecaption{Emission/Absorption Velocities of Principle Lines \label{tab:eclipse}}
\tablehead{
\colhead{Star} & \colhead{Line} & \colhead{RV (km/sec)} 
}
\startdata
AS 325 & H$\beta$ em & $\sim$0 \\
AS 325 & H$\beta$ ab & -61.7 \\
AS 325 & H$\alpha$ em & $\sim$0 \\
AS 325 & H$\alpha$ ab & -73.2 \\
AS 325 & NaI D em (5890) & $\sim$0 \\
AS 325 & NaI D ab (5890) & -76.3 \\
AS 325 & NaI D em (5896) & $\sim$0 \\
AS 325 & NaI D ab (5896) & -71.3 \\
XX Oph & H$\beta$ em & $\sim$0 \\
XX Oph & H$\alpha$ em & $\sim$0 \\
XX Oph & NaI D em (5890) & $\sim$0 \\
XX Oph & NaI D ab (5890) & -76.3 \\
XX Oph & NaI D em (5896) & $\sim$0 \\
XX Oph & NaI D ab (5896) & -76.4 \\
\enddata
\end{deluxetable}

\subsection{XX Oph}
\subsubsection{Long Term Photometry}

XX Oph has been photometrically monitored since 1890 starting with the Harvard plate collection \citep{Prager1940BHarO.912...17P,SW1926HarCi.292....1S}. Additional short term light curves have been presented as well in \citet{Merrill1932ApJ....75..133M,Evans1993A&A...267..161E, Cool2005PASP..117..462C, tarasoc2006ASPC..355..297T} plus the 30 years of continuous photometric coverage by the American Association of Variable Star Observers (AAVSO) providing visual and, more recently, multi-color CCD observations. 

Measurements taken from the Harvard observatory plate collection for XX Oph from 1890 to 1940 were presented in \citet{Prager1940BHarO.912...17P}. Starting in 1940, the AAVSO observers began to photometrically monitor XX Oph, and continue today, a total time period of over 31,000 days. 
The fifty years covered by the Harvard plate measurements presented in Prager show that XX Oph has a fairly stable maximum near m$_{pg}$ = 9.7, with five random sudden drops in brightness of 0.5-1.2 magnitudes covering varying time periods, the longest being $\sim$2 years. At this time, XX Oph was considered to be a member of the R CrB class of variables although the irregular brightness drops are much shallower than those seen in R CrB 
\citep[e.g.][]{RCB2013PASP..125..879H}.

\begin{figure*}
    \centering
\includegraphics[width=0.9\textwidth]{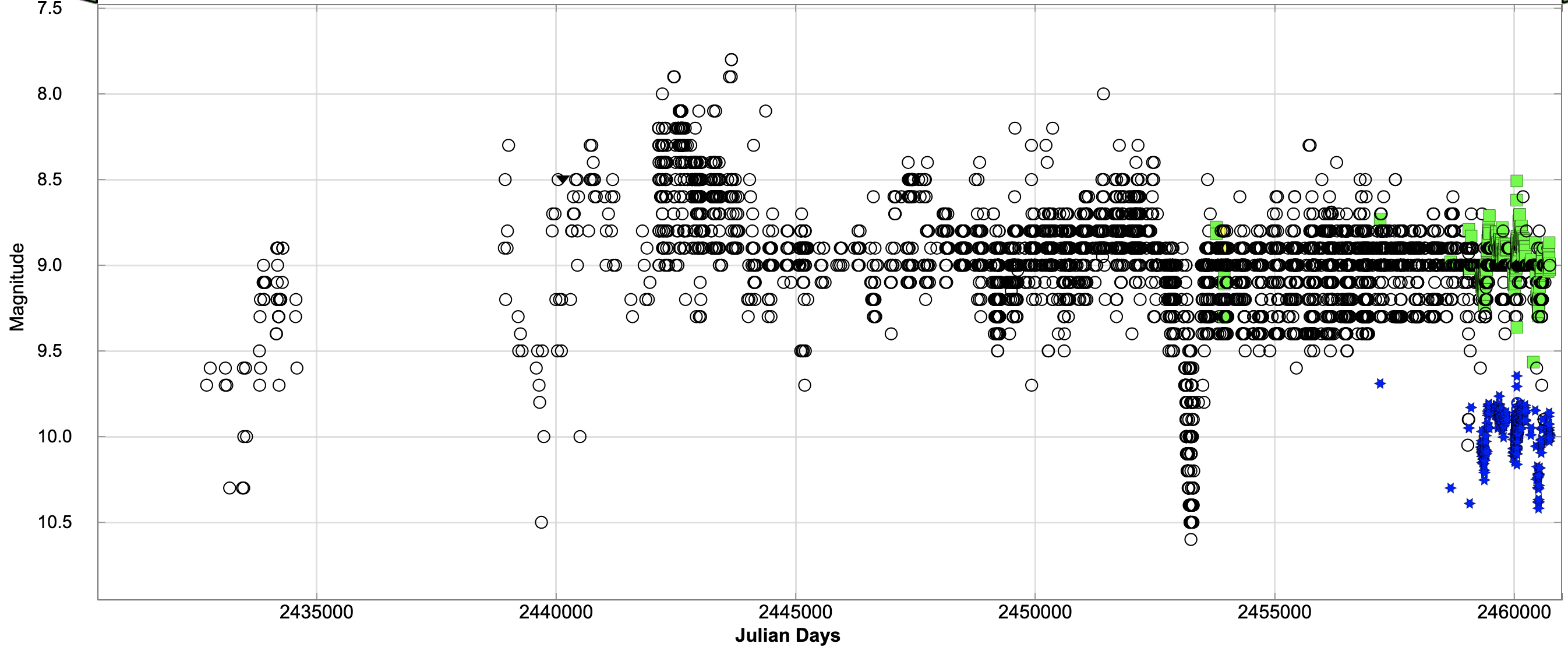}
\includegraphics[width=0.9\textwidth]{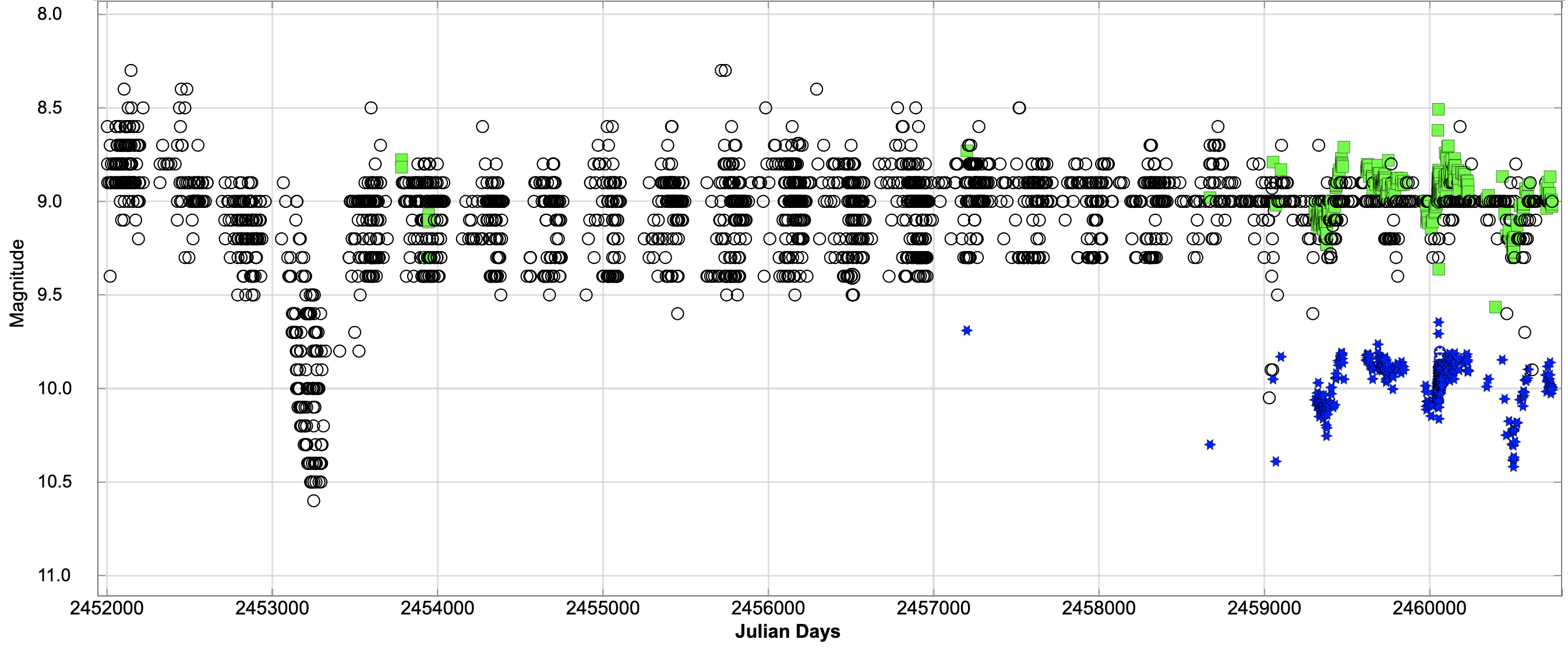}
\includegraphics[width=0.9\textwidth]{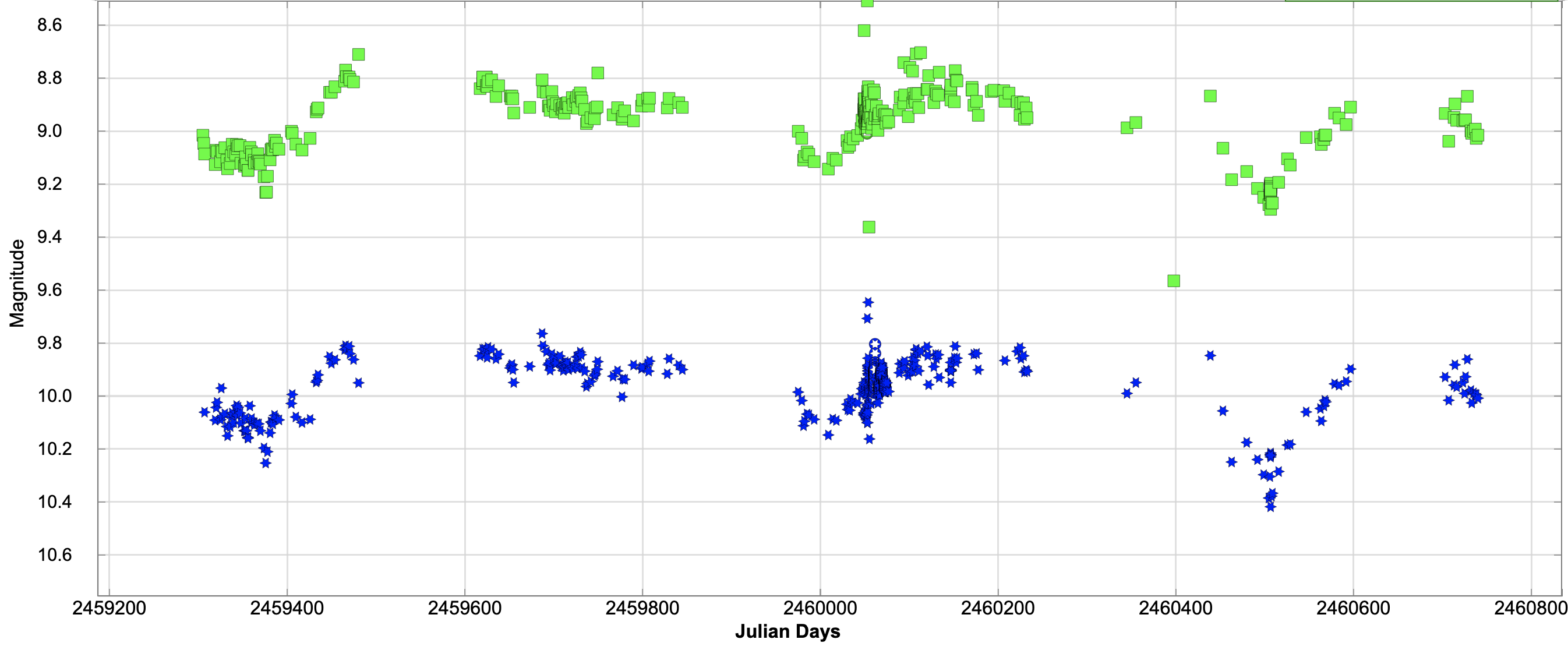}
\caption{AAVSO light curve of XX Oph covering a span of 85 years. 
Black points are visual measurements and blue and green symbols are CCD V and CCD B observations. The top panel (1941-2024) shows the earliest measurements with two or three large photometric dimmings visible. The middle panel (2001-2024) shows the time period from the presumed eclipse (see text) to the present, with no other deviations to near visual magnitude 10.5 observed. The bottom panel (2020-2024) shows recent CCD B and V measurements revealing a constant low-level scatter and a few small dips in the brightness of XX Oph. Data obtained from https://www.aavso.org/LCGv2/. }
\end{figure*}   

During the 85 years of AAVSO observations (Figure 7), only three deep fadings have been observed. Two poorly sampled fading events are seen (near JD 2433478 and JD 2439693) when XX Oph dropped to magnitude 10.4-10.5. The largest drop occurred near JD 2453250.3, when the star faded from its usual level at V=9.0 to V=10.6 for a few hundred days \citep{Cool2005PASP..117..462C,tarasoc2006ASPC..355..297T}. XX Oph has remained near visual magnitude $\sim$9 over most of its recorded light curve. 
The first two fading events are separated by 6214.3 days ($\sim$17 years). No other excursion to magnitude 10.5 is seen until the deep and wide fading event near JD 2453250, 13557.6 days later (approximately 2 times 6214 days). This few hundred day smooth decrease/increase event, had a shape reminiscent of a stellar eclipse \citep[as assumed in][]{Cool2005PASP..117..462C,tarasoc2006ASPC..355..297T}, and occurred at a time that is nearly twice the 17 year separation of the earlier dips. However, no fading event was seen at the half-way point or since JD 2453250.3. 

The bottom panel of Figure 7 shows the recent AAVSO Johnson B and V measures, revealing a variable pattern in the observations including some small dip-like features. These dips ($\Delta$B,V = 0.2-0.3 mag) produce no color change and may commonly occur in XX Oph, as evidenced by the scatter in visual magnitude (8.6-9.5) seen throughout the AAVSO light curve. Interestingly, the dip seen at JD 2459375.8 occurs 6126 days after the large dip seen at JD 2453250.3, a time period near the 6214 day spacing noted above. But the next dip seen, JD 2460500, is only 1100 days later.
Table 3 compares these five largest photometric fading events for XX Oph.
We can see that they do not seem to be similar in duration or depth.

\begin{deluxetable}{cccc}
\tablewidth{0pt}
\tablecaption{Major Photometric Fading Events of XX Oph\label{tab:fade}}
\tablehead{
\colhead{JD} & \colhead{Shape} & \colhead{Duration (days)} & \colhead{Depth (mags)} 
}
\startdata
2433478 & V? & 200? & 1.4 \\  
2439693 & V? & 800? & 1.5 \\
2453250.3 & U & 300 & 1.6 \\
2459375.8 & U? & 200 & 0.4 \\
2460500 & V & 150 & 0.5 \\
\enddata
\end{deluxetable}

\citet{tarasoc2006ASPC..355..297T} puts forth the idea that the JD 2453250 fading is indeed a stellar eclipse of the hot component by the cool giant star. More likely, however, is that evolved, late-type stars tend to be stochastically variable due to large photospheric convection cells and anomalous mass ejections \cite[e.g.,][]{Dupree2022ApJ...936...18D}, the events being 
one of a number of stochastic fadings caused by mass loss/ejection events coupled with variations of the M star \citep{Evans1993A&A...267..161E}. We know from \citet{lockwood1975ApJ...195..385L,Evans1993A&A...267..161E,Cool2005PASP..117..462C} that XX Oph is a binary star consisting of Be + M5 II-III stars similar to AS 325. 
If the large, wide drop to magnitude 10.5 was an eclipse, no other similar event has been seen in the 85 year light curve. As discussed above, there may be some small hint of dips (of various depths and widths) occurring on $\sim$6200 day cadences. If real, this may indicate an orbital modulation of about 17 years making XX Oph a very long period (eclipsing?) binary. More likely, however, is that evolved, late type stars tend to be stochastically variable due to large
photospheric convection cells and anomalous mass ejections \citep[e.g.,][]{Dupree2022ApJ...936...18D}  which create dust that can produce (deep) minima, similar to, but far less dramatic, than those seen in R CrB stars \citep[see the light curve in][]{Prager1940BHarO.912...17P}. Spectral evidence of such minor shell ejection events may occur as we discuss below. To date, no clear evidence exists for any coherent period or eclipse event in the light curve of XX Oph.

\begin{figure}
    \centering

\includegraphics[width=0.45\textwidth]{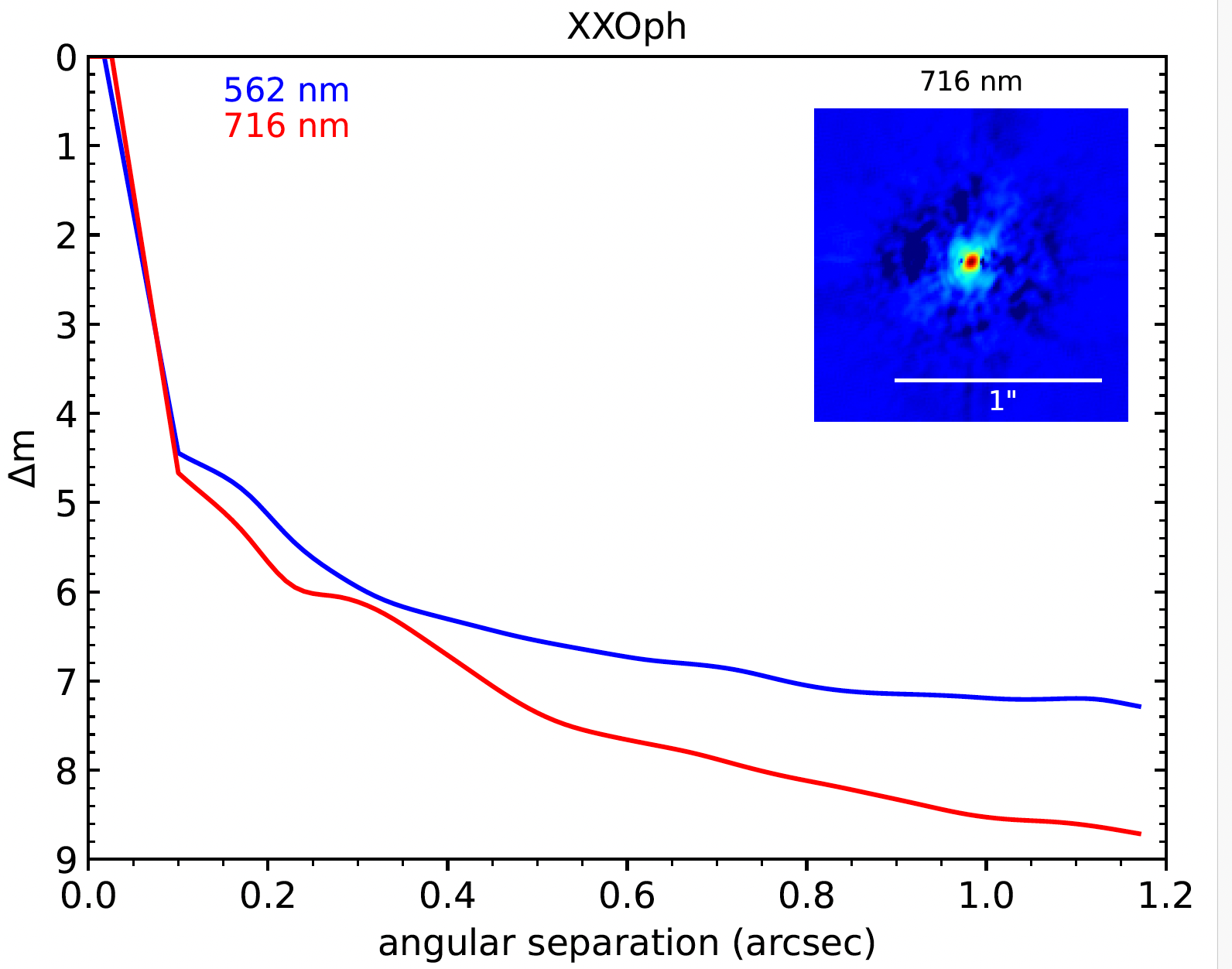}
\includegraphics[width=0.45\textwidth]{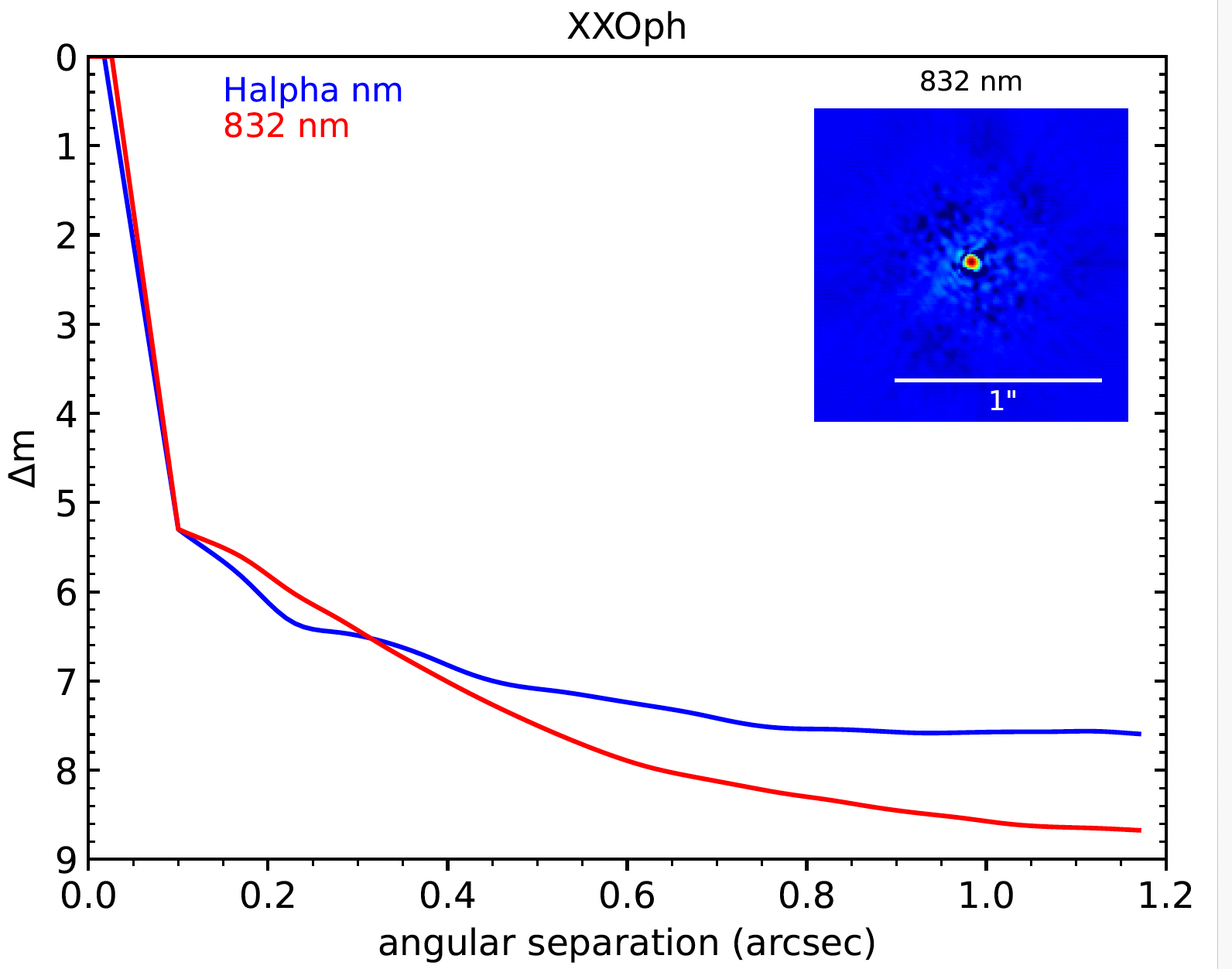}
\caption{Zorro high resolution speckle observation results for the star XX Oph obtained at Gemini South on 01 July 2023.  The red and blue lines show 5$\sigma$ magnitude contrast limits for the observations. Speckle reconstructed images are shown for 716 nm and 832 nm revealing extended nebulosity in the 716 nm band. No close stellar companions  are detected within 4-8 magnitudes from the diffraction limit out to 1.2 acrsec.}
\end{figure} 

\subsubsection{High-Resolution Imaging}

XX Oph was observed on 01 July 2023 using Zorro on the Gemini South 8-m telescope. Zorro \citep{scott2021FrASS...8..138S} obtains simultaneous speckle imaging in two optical bands and operates in a similar fashion as NESSI. For XX Oph, we obtained 25 sets of 1000 $\times$ 0.06 sec exposures collected and subjected to Fourier analysis in our standard reduction pipeline \citep{2011AJ....142...19H}. Figure 8 shows the 5$\sigma$ contrast curves obtained for XX Oph observed at 562 nm, 716 nm, H$\alpha$, and 832 nm. No close companion star to XX Oph is detected within the magnitude contrast achieved nor within the angular limits from the diffraction limit (20 mas) out to 1.2 arcsec, corresponding to spatial limits of 43 to 2571 au at the distance of XX Oph (d=2142.7 pc). At the distance of XX Oph, we did not, of course, expect to be able to separate the two stellar components. XX Oph sits within a dense H$\alpha$ cavity emitting reflected light from dust and gas and shows mid-IR PAH emission \citep{Howell2009PASP..121...16H}. The H$\alpha$ and 562 nm speckle data were of lower signal to noise and did not allow a robust reconstructed image to be produced. However, the 716 nm image shows extended emission surrounding the star in its dense environment. \citet{Evans1993A&A...267..161E} noted that the extinction to XX Oph mainly originates within the system itself and is quite high, A$_V$=1.6 magnitudes.

\subsubsection{Historical Spectroscopy}

XX Oph has a long history of spectral investigations. Starting with \citet{Merrill1924PASP...36..225M},
the odd "Iron Star" spectrum of XX Oph was noted. The nature of such strong, pure, and constant Fe II emission defied explanation as coming from a stellar photosphere. Bombardment by meteors was even suggested as the cause. \citet{Merrill1924PASP...36..225M,Merrill1932ApJ....75..133M,Merrill1951ApJ...114...37M,Merrill1961ApJ...133..503M} provide descriptions of the spectrum covering the years of 1921 to 1960 in which he noted that the bright emission lines of hydrogen, helium, and metals exhibit relatively small changes in shape and location while P Cygni lines appear at times in the spectrum (during mass ejections?) for species such as Na I D, hydrogen, helium, and the other metals. Extensive optical lines lists are provided in \citet{Merrill1951ApJ...114...37M,Merrill1961ApJ...133..503M,Cool2005PASP..117..462C}.
 
\citet{lockwood1975ApJ...195..385L} and \citet{deWinter1990Ap&SS.166...99D}
provide a look at the spectral energy distribution, IR spectroscopy, and a reddening measurement of XX Oph. These authors agree on the binary components (B0 III/V + M6-7 III) and conclude that much of an IR excess is due to extended shells of material, remnants of previous mass ejections.
\citet{Goswami2001BASI...29..295G,tarasoc2006ASPC..355..297T,TOM2010AJ....140.1758T} present high spectral resolution (R=20,000-60,000) snap shot views for XX Oph in which each find pronounced temporal variations in the absorption lines such as Na I D and hydrogen, especially in the width and depth of P Cygni profiles. Finally, \citet{Cool2005PASP..117..462C} and \citet{Howell2009PASP..121...16H} provide UV, optical, and IR IFU spectroscopy for XX Oph showing a possible correlation of P Cygni profiles appearing during the small non-periodic dips observed in the light curve (Figure 7). Over time, as shown in \citet{Cool2005PASP..117..462C, tarasoc2006ASPC..355..297T, Howell2009PASP..121...16H} when a light curve dip occurs, the absorption lines begin to appear, then deepen and broaden, and extend to higher blue shift until they fade. No evidence of long term (months to years) or short term (hours) coherent radial velocity changes indicative of binary motion have been detected \citep{Cool2005PASP..117..462C}.

\subsubsection{GHOST spectroscopy}

\begin{figure*}
\includegraphics[width=0.95\textwidth]{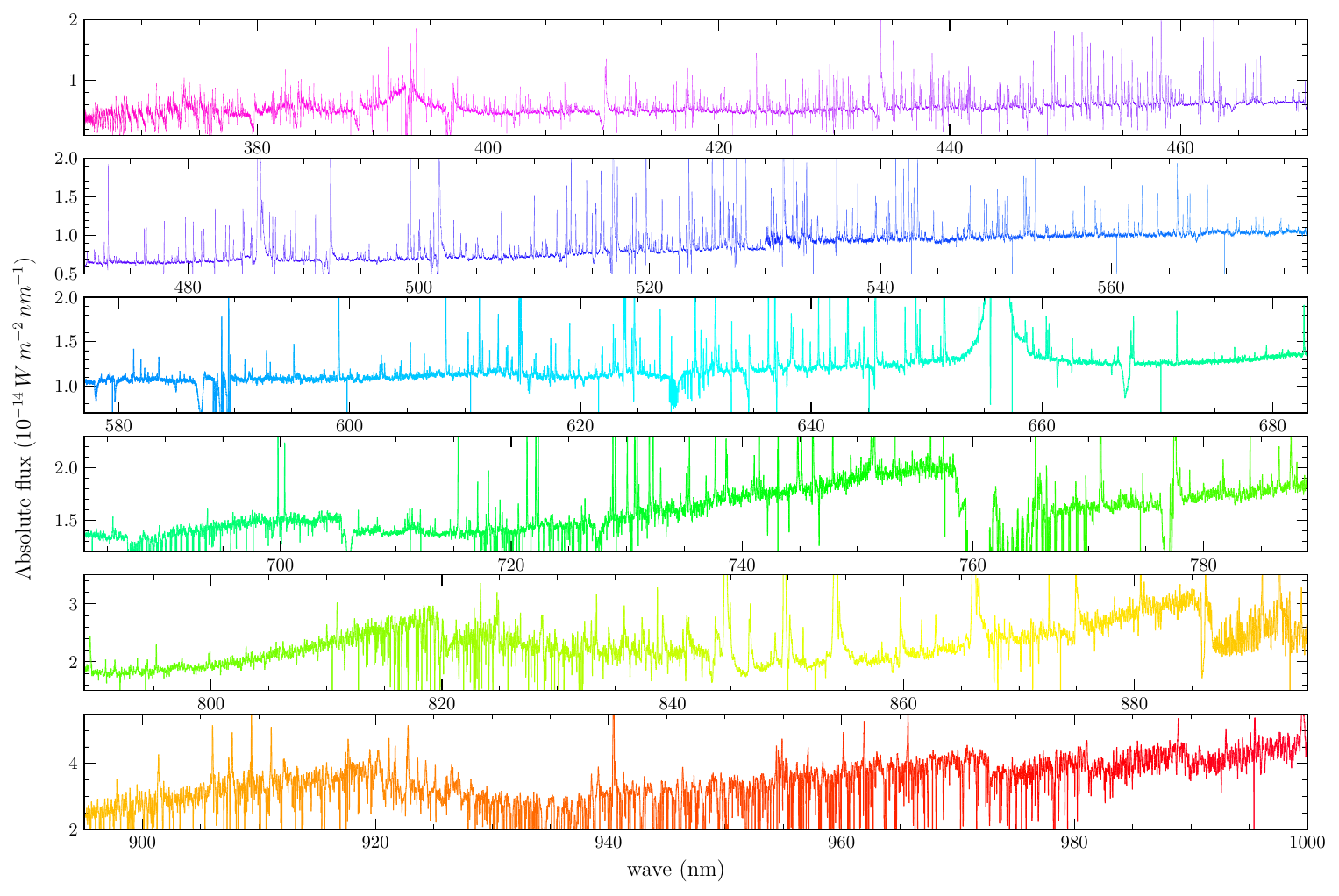}
\caption{Flux calibrated high resolution GHOST spectrum of XX Oph obtained on 17 April 2023. The various spectral orders are color coded for convenience showing the complex spectrum from 370-1000 nm.\label{xspectra}}
\end{figure*}

The full optical flux-calibrated GHOST spectrum of XX Oph is shown in Fig.\,\ref{xspectra}. Data were taken on the same night as AS 325, with the same exposure times and instrumental settings, but in high-resolution mode achieving a spectral resolution of 76,000. Data were reduced and analyzed in the same manner.  

We note that TiO molecular bands are clearly present redward of 700 nm coming from the M2 II-III star similar to those noted in \citet{Cool2005PASP..117..462C}.  We also identify multiple diffuse interstellar bands (DIBs), all suggesting a high extinction towards the source (E(B-V)$\sim$1\,mag following the relationships presented in \citealt{2019ApJ...878..151F}), agreeing with previous determinations using spectroscopic parallax (e.g.,\citealt{Evans1993A&A...267..161E}). Well-known strong DIBs seen include 5780, 5797, 6284\AA~ (affected by tellurics), but also lesser known, weaker DIBs. 

\begin{figure*}
\centering
\includegraphics[width=0.45\textwidth]{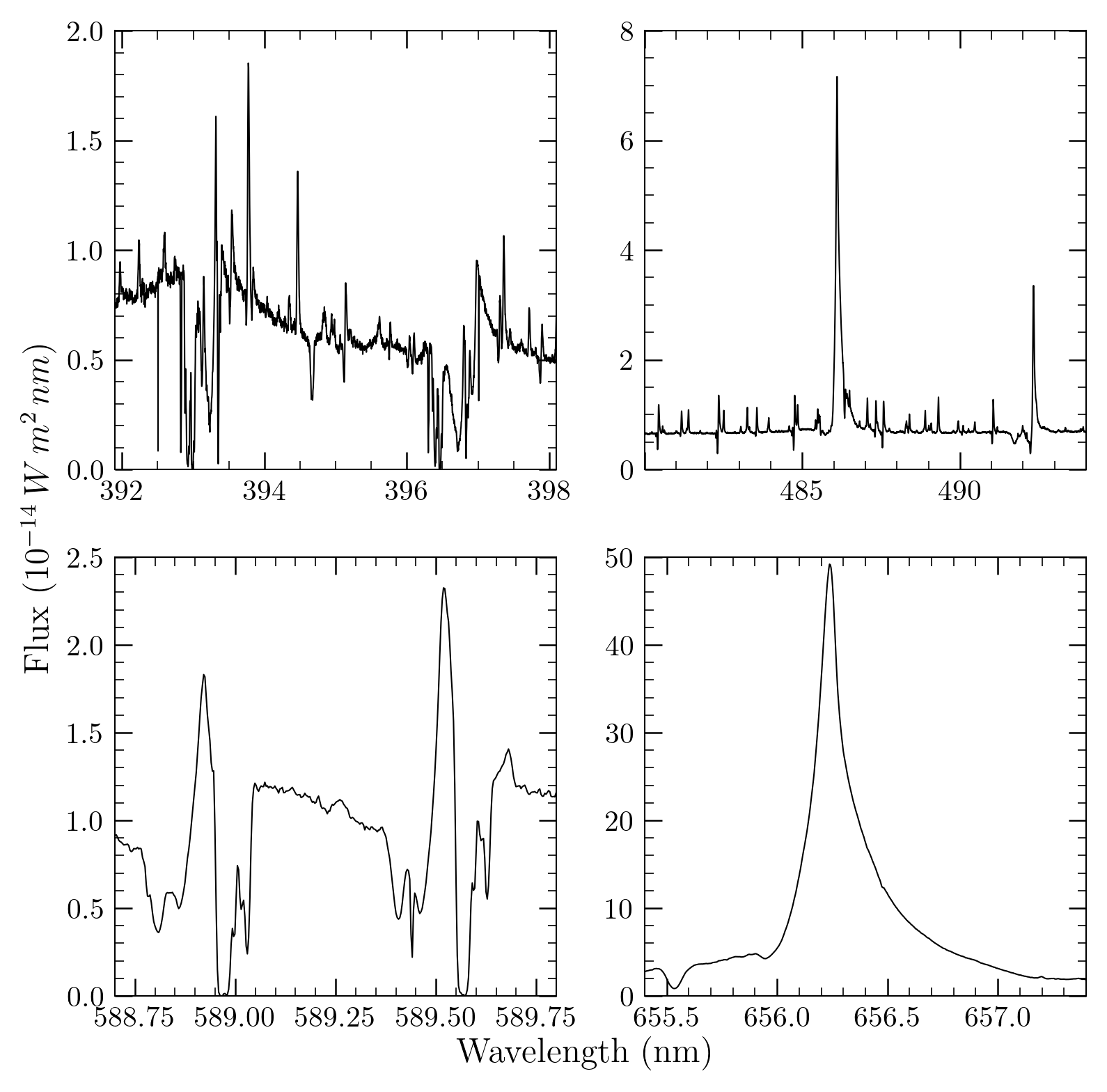}
\includegraphics[width=0.45\textwidth]{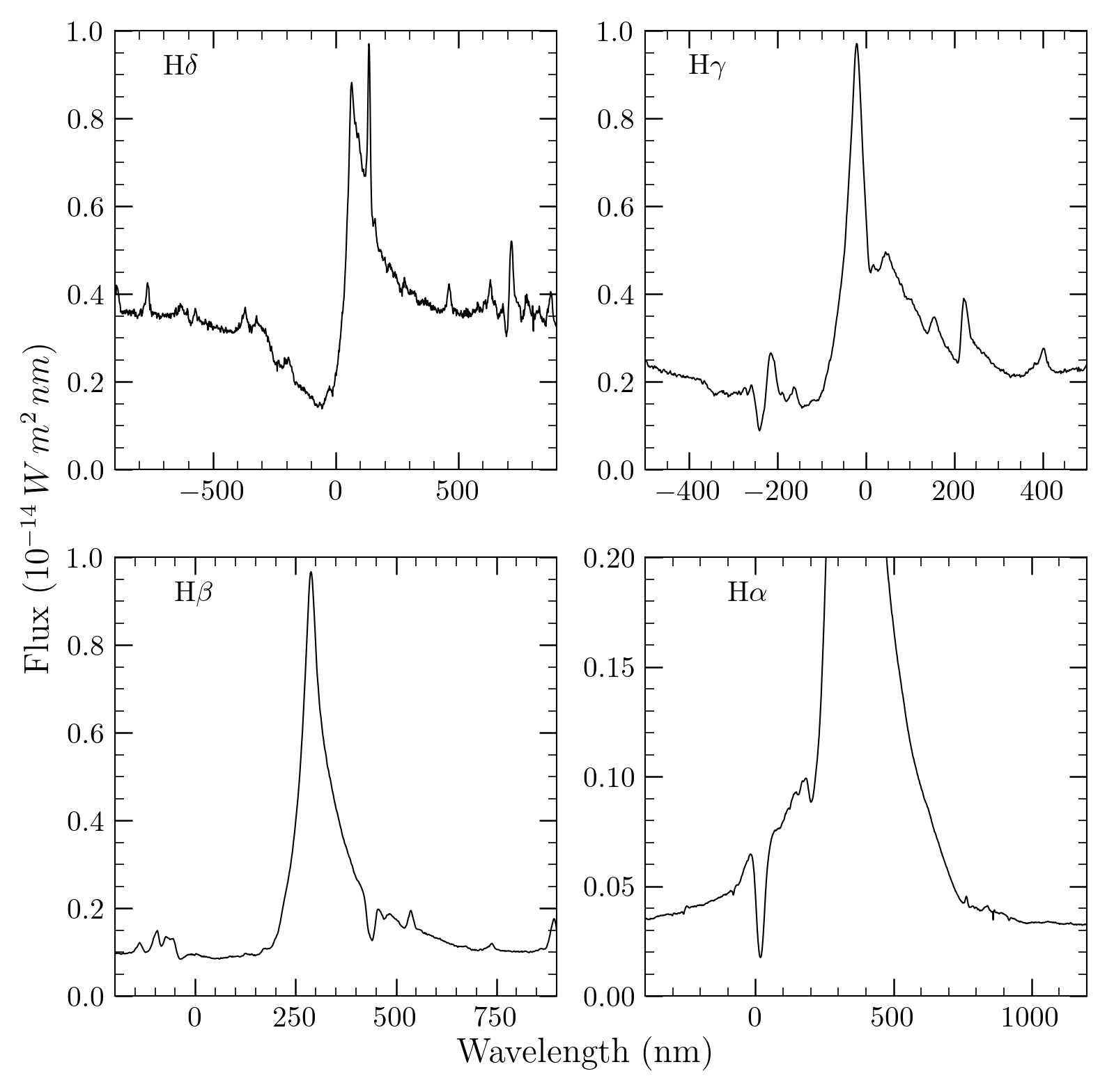}
\caption{Zoomed in selected spectral regions of interest for XX Oph covering Ca II H\&K, H$\beta$, NaI D, H$\alpha$ (Panel a), with select Balmer lines shown in (Panel b).
\label{xzooms}}
\centering
\end{figure*}

The forest of emission lines is impressive with a variety of P Cygni profiles visible. The light curve behavior at this time (JD=2460052.3), seen in Figure 7, shows that XX Oph was $\sim$50 days past the most recent small light curve dip. Emission lines in XX Oph tend to exhibit only small changes in intensity and velocity. This was noted by \citet[][and references therein]{Merrill1961ApJ...133..503M} and others with the emission lines being near $-$40 km\,sec$^{-1}$ at all epochs. 
The spectrum shows strong emission lines prominently of many ionized metals, including Ti, Mg, O, Na, Fe, and Ca, with lines towards the blue more often visible with absorption components. There are also weak forbidden emission lines in [FeII], [SII] 4068\AA, [OI] 6300\AA, and [CaII] 7291\AA. He lines are present in the spectra, mostly in absorption or in P Cygni. Prominent He\,I emission lines at 6678\AA\ (three components), and 4471\AA, suggest a hot star (i.e. a mid-B spectral type or later as no He\,II was identified). Multiple Ca\,II lines are visible including the near-infrared triplet (multiple peaks in emission), and deep complex Ca\,II H and K lines with multiple troughs suggestive of a late M-type component. Notable are the unique Na I D lines, with broad absorption components (see Fig.\,\ref{xzooms}). The presence of both P Cygni and inverse P Cygni suggest possible variable mass flows from the central hot star. We also draw attention to the changing line profiles of the Balmer lines, which are indicative of the binary nature of the XX\,Oph system. In the observed spectrum, deep asymmetric blue shifted absorption is clearly visible in H$\delta$ (and with lower SNR at higher order Balmer lines), but this vanishes towards lower level transitions. In H$\beta$ and H$\alpha$ there appears no noticeable absorption. 

The P Cygni absorption lines, likely due to mass ejections, tend to behave erratically with blue shifted velocities of $-$50 to $-$370 km\,sec$^{-1}$ seen at times \citep{Merrill1961ApJ...133..503M,Cool2005PASP..117..462C,TOM2010AJ....140.1758T}. As noted above, there is likely a connection between light curve dips (mass ejections?) and the appearence of P Cygni lines. While it is clear that XX Oph is a binary system, it does not eclipse thus, we do not know the orbital phase at the time of the GHOST spectrum. During this epoch, the Balmer lines have very asymmetric profiles. The Ca II H\&K lines and the Na I D lines are complex with multiple components, some of which are likely to be local ``ISM" lines from the dense medium surrounding XX Oph. The Na I D lines reveal complex structures similar to those seen by \citet{TOM2010AJ....140.1758T}.
The NaI D lines show blue-shifted and red-shifted absorption components, some possibly due to line-of-sight/local ISM material.
Table 2 lists the measured values for the lines shown in Figure 10.

\section{Discussion}

\subsection{AS 325}
\subsubsection{Light Curve}

The full AAVSO light curve, Figure 1, presents one very deep and wide eclipse (eclipse 9). This eclipse occurred during a time when the red secondary star presented spectral features consistent with an M5 III star \citep{Howell2009PASP..121...16H} while during other eclipses with contemporaneous spectra \citep{Cool2005PASP..117..462C} the secondary star has a spectral type of K2.5 III. The eclipse duration tends to remain near 90-100 days while the depth typically is near 0.8-1.0 magnitudes. Some times in the light curve, for example near eclipses 15 \& 16, any eclipse is hard to detect due to a chaotic nature of the light curve.  
Stellar models for K2.5 II and M5 II stars \citep{2000asqu.book.....C} suggest these stars have radii of 200 and 500 solar radii, a factor of 2.5, approximately the same value as the ratio of eclipse duration between the longest (eclipse 9, M5 II secondary) and typical eclipses (K2.5 II secondary). 

The distance to AS 325 was estimated to be 1100 pc by \cite{Cool2005PASP..117..462C} and $>$1000 pc by \citet{Howell2009PASP..121...16H}, each based on a spectroscopic parallax estimate assuming the secondary star to dominate the light near V band and having a luminosity class near II-III. Gaia lists the distance to AS 325 as 5339 pc, a value which fits well for a luminosity class II or III secondary star with spectral type near M2 to M5. 

\subsubsection{Orbital Period}

\citet{Otero2005IBVS.5608....1O} and \citet{Howell2009PASP..121...16H}
both found that the orbital period of AS 325 was 513 days, based on light curve eclipse mid-times. With the extended light curve presented in this paper, we decided to reexamine the eclipse times and independently establish the orbital period for AS 325.

Given the eclipse mid-times listed in Table 1, we can, in principle, measure the time between each pair of eclipses and arrive at a good estimate of the orbital period for AS 325. However, since many of the eclipses are not well sampled, having uncertainties of 5-10 days for some in terms of the mid-eclipse, we made use of eclipse pairs with good measurements for the mid-eclipse time (e.g., eclipse pairs 4-5, 5-6, 9-11, 14-22 with pairs 7-8 and 9-11 spanning two eclipses in time). Taking these eclipse pairs mid-times and averaging the resulting values, we find that the orbital period for AS 325 is 512.943$\pm$1.06 days. We give the mid-eclipse phase 0.0 ephemeris below.

$$
Phase_{mideclipse}HJD=2460238.554682 + 512^d.943\pm1^d.06~x~E
$$

\subsubsection{Spectral Properties}

\citet{Howell2009PASP..121...16H} noted TiO bands from the secondary star and, along with IR spectra presented in the same paper, determined the secondary star to be of spectral type M5 II. These spectral observations were obtained near in time to the longest and deepest AS 325 eclipse ever seen (See Table 1, eclipse 9). \citet{Cool2005PASP..117..462C}  and the GHOST spectrum presented here show no TiO bands, the recent eclipses being shorter and shallower (See Figure 3) and spectral features from the secondary star are more consistent with a spectral type of K2.5 II. The pulsational evolution of the secondary star in AS 325 is thus likely to be on the order of decades or longer.

AS 325 has a listed distance, using the Gaia parallax, of 5339 pc. \citet{Cool2005PASP..117..462C} quoted a distance of 1100 pc based on matching their red spectra to an K2.5 II secondary while \citet{Howell2009PASP..121...16H} gave a lower limit to the distance of $>$1000 pc based on their optical red spectral match to a M5 II star. To match the now known Gaia distance and visual magnitude with a spectroscopic parallax, the red optical light would be dominated by a late K or early M luminosity class II or Ib star. This secondary star type is in good agreement with the previous estimated spectral and luminosity class for the secondary star.

\subsection {XX Oph}

XX Oph, at a Gaia distance of 2143 pc, a visual magnitude of 9.0, and adopting the reddening value determined by \citet{Evans1993A&A...267..161E} (A$_V$=1.6), will have its red optical continuum dominated by a luminosity class M6 II star of M$_v$$\sim$-4.3 \citep{2000asqu.book.....C}. 

\citet{lockwood1975ApJ...195..385L,Evans1993A&A...267..161E,Cool2005PASP..117..462C} all concluded that the secondary star in XX Oph would be near M6-7 III but each of these authors adopted a distance to XX Oph of near 1900 pc, about 20\% less than the current Gaia determined value.   
\citet{Goswami2001BASI...29..295G} 2001 obtained two R=60,000 spectra about one year apart, revealing dramatic changes in the P Cygni profiles, especially for NaI D. \citet{TOM2010AJ....140.1758T}, using a R=20,000 spectrum obtained near in time to that of \citet{Goswami2001BASI...29..295G}, also noted the large P Cygni profiles present in XX Oph, He I for example, showing a radial velocity of -358 km/sec. Our spectrum (R=76,000), taken 22 years later, shows generally similar hydrogen, He, and NaI D line profiles as seen in Fig.\,\ref{xzooms}.

\subsection{Nature and Evolution of the Systems}

\begin{deluxetable*}{ccc}
\tablewidth{0pt}
\tabletypesize{\scriptsize}
\tablecaption{Properties of AS 325 and XX Oph\label{tab:fade}}
\tablehead{
\colhead{Property} & \colhead{AS 325} & \colhead{XX Oph}
}
\startdata
Distance (pc) & 5339 & 2143 \\
Environment & Dense Galactic-plane region with extended H$\alpha$ emission & Dense Galactic-plane region with extended H$\alpha$ emission \\
Component Stars & BeV + K2.5-M5III (variable) & Mid-Be III/V + M6 III \\
Orbital Period (days) & 513 & unknown \\
Eclipse? & Yes & No \\
Photometric Variability & Eclipses up to 2 mag deep & Irregular drops up to 1.5 mag deep \\ 
Normal Magnitude (m$_V$) & 10.0 & 9.6 \\
\enddata
\end{deluxetable*}

Table 4 provides a comparison of the properties for the two Iron Stars. They share similar stellar components and show photometric variability. The orbital period of XX Oph is yet to be determined.

Both iron stars started their life as a pair of massive main sequence stars. they currently consist of Be primaries, with late-type close pulsationally variable bright giant companions. Such systems, with fairly extreme mass ratios $<\sim$0.1 are not very common and are often broadly lumped together in the $\zeta$ Aurigae class \citep{2015ASSL..408.....A}. While no other $\zeta$ Aurigae stars show the spectacular array of metallic emission lines seen in AS 325 and XX Oph, due to their current evolutionary status, the stars in the class are long period, (atmospheric) eclipsing systems and are also famous stars, e.g., 
$\epsilon$ Aurigae and $\eta$ Geminorum. The stars discussed herein form an important example within the class in which one can study the interaction of colliding winds; a low-density, radiation-driven wind from the hot primary star and a low-velocity, clumpy wind from the low-mass cool giant secondary. The complex and varied spectral line profiles seen in both the Balmer lines and the ionized metals reflect a complex circumstellar environment. 

Long term, detailed photometric observations are on-going. Contemporaneous high resolution spectral studies of these two Iron Stars would allow a wide array of astrophysics to be gleaned for both of these unique binaries.
Temporal changes in the stars, their winds, their wind interactions, and outflows could be measured during these stars' astronomically fast evolutionary times. 
Given their brightness, such a program is quite feasible, albeit one that would be years in the making. We note that the raw GHOST spectra for AS 325 and XX Oph are available in the Gemini Observatory data archive. The fully reduced spectra for both stars, used herein, are available as on-line only tables.


\section*{Acknowledgments}
We wish to thank the many observers of the AAVSO for their dedication and multi-year photometric observations of the stars AS 325 and XX Oph.
Some of the observations in this paper made use of the High-Resolution Imaging instrument Zorro and were obtained under Gemini LLP Proposal Number: GN/S-2021A-LP-105. Zorro was funded by the NASA Exoplanet Exploration Program and built at the NASA Ames Research Center by Steve B. Howell, Nic Scott, Elliott P. Horch, and Emmett Quigley. GHOST was built by a collaboration between Australian Astronomical Optics at Macquarie University, National Research Council Herzberg of Canada, and the Australian National University. GHOST/Zorro are mounted on the Gemini South telescope of the international Gemini Observatory, a program of NSF’s NOIRLab, which is managed by the Association of Universities for Research in Astronomy (AURA) under a cooperative agreement with the National Science Foundation, on behalf of the Gemini partnership: the National Science Foundation (United States), National Research Council (Canada), Agencia Nacional de Investigación y Desarrollo (Chile), Ministerio de Ciencia, Tecnología e Innovación (Argentina), Ministério da Ciência, Tecnologia, Inovações e Comunicações (Brazil), and Korea Astronomy and Space Science Institute (Republic of Korea).

\bigskip

\noindent {\it {Facility:}} Gemini:South, Zorro, GHOST; WIYN:NESSI

\bibliographystyle{aasjournal.bst}
\bibliography{AS325.bib}








\end{document}